\documentclass[prd,aps,superscriptaddress,showpacs,amssymb,twocolumn]{revtex4}
\usepackage{graphicx,dcolumn}
\def\bacol{\setlength{\arraycolsep}{0pt}}

\newcolumntype{e}[1]{D{.}{\cdot}{#1}}

\begin{document}

\title{The contribution of ${\cal O}(\alpha)$ radiative corrections to
the renormalised anisotropy and application to general tadpole
improvement schemes: addendum to ``One--loop calculation of the
renormalised anisotropy for improved anisotropic gluon actions on a
lattice''}

\author{I.T. \surname{Drummond}}
\affiliation{DAMTP, CMS, University of Cambridge, Wilberforce Road,
Cambridge CB3 0WA, U.K.}
\author{A. \surname{Hart}}
\affiliation{School of Physics, University of
Edinburgh, King's Buildings, Edinburgh EH9 3JZ, U.K.}

\author{R.R. \surname{Horgan}}
\author{L.C. \surname{Storoni}}
\affiliation{DAMTP, CMS, University of Cambridge, Wilberforce Road,
Cambridge CB3 0WA, U.K.}

\begin{abstract}
General ${\cal O}(\alpha)$ radiative corrections to lattice actions
may be interpreted as counterterms that give additive contributions to
the one--loop renormalisation of the anisotropy. The effect of
changing the radiative coefficients is thus easily calculable.  In
particular, the results obtained in a previous paper for Landau mean
link improved actions apply in any tadpole improvement scheme. We
explain how this method can be exploited when tuning radiatively
improved actions. Efficient methods for self--consistently tuning
tadpole improvement factors are also discussed.
\end{abstract}

\preprint{Edinburgh 2003/09}


\maketitle

\section{Introduction}
\label{sec_introduction}

The use of anisotropy is becoming an increasingly important and
prevalent tool for improving lattice simulations. This stems from its
utility in increasing the temporal or momentum resolution of
correlation functions without the associated computational burden of
reducing the lattice spacing in all directions.

Its use is complicated by the multiplicative renormalisation that
occurs on the lattice. In a previous paper
\cite{Drummond:2002yg}
we argued that one--loop lattice perturbation theory can be used to
calculate the renormalisation. We further demonstrated that a one--loop
calculation of the renormalisation is sufficient to describe the
measured value to an accuracy of around 3\%. Such a determination is
more than sufficient for most lattice calculations, including
spectroscopy.

These calculations were carried out for the Wilson gauge action, and
for a Symanzik improved action also. Results were presented for
theories with and without tadpole improvement.  The tadpole factors
used were from mean links measured in Landau gauge. We have since
received enquiries about the anisotropy renormalisation for other
definitions of tadpole improvement. Our work does, in fact, cover such
situations, but requires a slightly different presentation to make
this clear.

In this note we clarify this issue. We illustrate the general
applicability of our results by calculating the anisotropy for the
aforementioned actions using generic tadpole improvement schemes,
including the mean plaquette. We discuss also how this approach may be
extended to actions further modified with, for instance, radiatively
corrected improvement factors.

\section{Method}

For clarity, we briefly summarise here our method
\cite{Drummond:2002yg}.
We require a quantity that can be described perturbatively, yet which is
sensitive to the anisotropy. Using a 4--dimensional lattice twisted in
the $1,2$--directions to provide a gauge invariant gluon mass
\cite{luwe},
we studied the infrared dispersion relation of the on--shell gluon in
Feynman gauge. We consider the gluon mode (known as the $A$ meson in
\cite{luwe})
which has momentum (Euclidean, and in the isotropic frame)
\begin{equation}
\mathbf{p} = (iE_0,m_0,0,p_3), \mbox{\hspace{3em}}
m_0 = \frac{2 \pi}{NL} \; .
\end{equation}
$N$ is the number of colours, and $L$ the extent of the lattice
in the twisted directions.

We restrict our discussion here to the case where the lattice spacing
is smaller in the temporal direction only, by a factor of $\chi$. The
on--shell condition for very small momentum in the (untwisted)
3--direction reads
\begin{equation}
\frac{1}{\chi} \left[
-\chi^2 F_t(E_0) + p_3^2 + F_s(m_0) \right] = 0 \; .
\label{eqn_disp_reln}
\end{equation}
For the Wilson action, $F_s(m) \equiv 4 \, \sin^2(m/2)$, and more
complicated functions with the same continuum limit for (Symanzik)
improved cases. For actions unimproved in the temporal direction,
$F_t(E) \equiv 4 \sinh^2 (E/2)$.

The one--loop contribution to the self energy, $\Sigma$, was
calculated using lattice perturbation theory. This leads to a
corresponding renormalisation of the anisotropy, $\chi_R = \chi (1 +
\eta g^2)$, and of the twisted mass, $m_R = m_0 + m_1 g^2$. These can
be extracted by looking at the variation of the $\mu=\nu=2$ component
of the self energy with $p_3$:
\begin{equation}
2 \eta \left[ p_3^2 + F_s(m_0) \right] - m_1 F^\prime_s(m_0) = 
\chi \Sigma_{22}(\mathbf{p}) \; .
\end{equation}
There are two important points to note:

\begin{enumerate}
\item
For the one--loop calculation, the on--shell condition is that derived
from the tree level inverse propagator. The addition of radiative
terms to the action, \textit{i.e.} those of ${\cal O}(g^2)$ and above,
does not change Eqn.~(\ref{eqn_disp_reln}).

\item
The renormalisation of the anisotropy, $\eta$, is proportional to the
terms in the one--loop self energy, $\Sigma$, that vary as $p_3^2$. 
Radiative terms of ${\cal O}(g^2)$ give counter--term insertions in the 
gluon propagator that yield additive corrections to $\Sigma$, and hence 
additive corrections to $\eta$.

\end{enumerate}

The counter--terms themselves consist of a coefficient, $d$, that is
separately determined and a 2--point vertex function arising from the
perturbative expansion of a combination of closed contours of links in
the action. The contribution of the latter to the self energy,
$\Sigma_{\text{ct}}$, leads to a change in the anisotropy
renormalisation
\begin{equation}
\delta \eta = \eta_{\text{ct}} d \; .
\end{equation}
It is clear that the results can be applied to a whole class of
actions differing only in the choice of $d$. The obvious example is
for choices of improvement factor in a tadpole improved action. Another
interesting case is radiatively improved theories, where we might wish
to change the criterion for improvement, and hence the ${\cal O}(g^2)$
couplings in the action.

In all these examples, the anisotropy does not need to be recalculated
for a change in $d$ if have $\eta_{\text{ct}}$. This is an obvious
benefit when we are, say, in the process of tuning radiative corrections
to improved actions.

\section{Results}

In this section we present $\eta_{\text{ct}}$ for a Symanzik and
tadpole improved action
\cite{Drummond:2002yg}.
The original action contains parameters $\beta_0$, $\chi_0$ and one
tadpole improvement factor, $u_{s,t}$, per spatial or temporal link
respectively. With a rescaling
\begin{equation}
\beta = \frac{\beta_0}{u_s^3 u_t} \; , \,\,\,  
\chi = \chi_0 \frac{u_s}{u_t} \, , 
\label{eqn_rescaling}
\end{equation}
we obtain
\bacol{
\begin{eqnarray}
S_{SI}&&(\beta,\chi,u_s) ~ = ~
-\beta \chi \sum_{x,s} \left\{
\frac{4}{3} P_{s,t} - \frac{1}{12}\frac{R_{ss,t}}{u_s^2} \right\}
\nonumber \\
 -\beta &&\frac{1}{\chi} \sum_{x,s>s'} \left\{
\frac{5}{3} P_{s,s'} - \frac{1}{12}\frac{R_{ss,s'}}{u_s^2}
-\frac{1}{12}\frac{R_{s's',s}}{u_s^2} \right\},
\label{eqn_si_act}
\end{eqnarray}
}
where $P$, $R$ are plaquettes and $2 \times 1$ loops respectively, $t$
denotes the temporal direction, and $s$, $s'$ run over the spatial
directions.

If the spatial tadpole factor may be written as $u_s = 1 + d_s g^2 + {\cal
O}(g^4)$, then the action can be expressed as a ``bare'' action, plus a
counter--term: 
\begin{equation}
S_{SI}(\beta,\chi,u_s) = S_{SI}(\beta,\chi,u_s=1) +
g^2\Delta S_{SI} + {\cal O}(g^4) \; ,
\end{equation}
where
\bacol{
\begin{eqnarray}
\Delta S_{SI}  =&&  - \beta d_s \sum_{x,s>s'} - \frac{1}{6} \left\{
\chi R_{ss,t} + \frac{1}{\chi} \left( R_{ss,s'}+R_{s's',s} \right) \right \}
\nonumber \\ 
=&& \beta S_{\text{ct}} \; .
\end{eqnarray}
}
We can see at this point what the result for $\eta_{\text{ct}}$ will
be by looking at the terms in the action that will contribute via
$\Sigma_{22}$, \textit{viz.} terms in $A_2 p_3^2 A_2$. It is clear
from Eqn.~(\ref{eqn_si_act}) that to obtain the correct continuum
limit $R_{ss,t}$ must yield a term proportional to $4 \chi A_2 p_0^2
A_2$ (using continuum notation), and $R_{ss,s'}$, $R_{s's',s}$ both
proportional to $4 A_2 p_3^2 A_2/\chi$. For an on--shell propagator,
$\chi^2 p_0^2 + p_3^2 = 0$. Hence the Symanzik improvement $\chi^2
R_{ss,t} + R_{ss,s'} + R_{s's',s} \propto 4 A_2 p_3^2 A_2$. The
counter--term $\Delta S_{SI}$ hence contributes $\Sigma_{\text{ct}} =
-4/(6 \chi)$, or
\begin{equation}
\delta \eta(\chi) = - \frac{d_s(\chi)}{3}   .
\label{eqn_delta_eta}
\end{equation}
We have numerically checked this. We expand $S_{\text{ct}}$ as a sum
of vertex functions using a {\sc Python} code
\cite{Horgan_in_prep},
and calculate the contribution to the one--loop self energy by summing
over lattice modes for lattices of extent $I=50$ in the untwisted
directions. The twisted extent was varied in the range $4 \le L \le
32$. The corresponding contribution to the anisotropy is extrapolated
$L \to \infty$ as per
\cite{Drummond:2002yg}.
The value for $\eta_{\text{ct}}$ is accurate to the rounding error
(the $6^{\text{th}}$ decimal place), and has been calculated over a
range of anisotropies $1 \le \chi \le 8$.

As a further check, we test these results for the mean link in Landau
gauge.  The renormalisation of the anisotropy for the action without
tadpole improvement is (Eqn.~(98) of
\cite{Drummond:2002yg})
\begin{equation}
\eta(\chi) = 0.0602 - \frac{0.0656}{\chi} - \frac{0.0237}{\chi^2} \; .
\label{eqn_eta}
\end{equation}
The one--loop expression for the mean link in this gauge is 
\cite{Drummond:2002kp,Drummond_in_prep}
\begin{equation}
d_s(\chi) = -0.1012 + \frac{0.0895}{\chi} - \frac{0.0513}{\chi^2} -
\frac{0.0502 \log \chi}{\chi} \; .
\label{eqn_d_s}
\end{equation}
Combining
Eqns.~(\ref{eqn_delta_eta}),~(\ref{eqn_eta}),~(\ref{eqn_d_s}) yields
the mean link improved result with a fit formula consistent with that
given in Eqn.~(99) of \cite{Drummond:2002yg}.

\section{Application}

An efficient procedure for carrying out a lattice calculation is thus
as follows. It is more convenient to begin with the rescaled action,
Eqn.~(\ref{eqn_si_act}), as there is no reference to $u_t$ (there
being no Symanzik improvement in the temporal direction). Having
selected $\beta$ and $\chi$, we vary $u_s$ until we have a
self--consistent value (\textit{i.e.} the measured expectation value
$\langle u_s \rangle$ in the chosen tadpole improvement scheme agrees
with $u_s$ used in the action). We now measure the expectation value
of $u_t$, and \textit{define} its self--consistency through $\beta_0 =
\beta u_s^3 u_t$ and $\chi_0 = \chi u_t/u_s$. The tadpole improved
coupling is $g_0^2 \equiv 6/\beta_0$.

The one--loop renormalisation of the anisotropy, $\eta(\chi)$ is
calculated using the formulae in the previous section. The actual
ratio of length scales in different directions is then given to
one--loop by
\begin{equation}
\chi_R = \chi(1+\eta(\chi) g_0^2) \; .
\label{chi_R}
\end{equation}
For Landau gauge mean link tadpole scheme this prediction is accurate
to within 3\%
\cite{Drummond:2002yg}.

The self--consistent values of $u_s$ and $u_t$ can alternatively be
calculated in perturbation theory using the chosen tadpole improvement
scheme
\cite{Drummond:2002kp}.
It is a tenet of tadpole improvement, however, that it is better to
demand numerical self--consistency of $u_{s,t}$, rather than agreement
of the perturbative expansion coefficients to a given order. We are
then led to ask whether a similar ``non--perturbative'' definition
$d_s = (1-u_s^{-2})/(2g_0^2)$ (Eqn.~(\ref{eqn_si_act}) suggests this
form) might improve the anisotropy calculation. We do not find this to
be the case; the resulting $\chi_R$ over-estimates the measured value
by around 6\%.

It is apparent now from where the dependence of $\chi_R$ on the choice
of tadpole improvement arises.  The variation of the one--loop factor
$d_s$ between plaquette and Landau mean link improvement schemes is
small. Thus for given $\chi$, $\eta$ shows little variation with the
choice of scheme.

The tadpole improvement parameters for these two schemes are
different, however, but the difference is only apparent from
simulation results or in two and higher loop perturbation theory. For
this reason, if we are simulating the action with chosen parameters
$(\beta_0,\chi_0,u_{s,t})$, the value of the rescaled $\chi$ is
tadpole improvement scheme dependent. The result is that the majority
of the scheme dependence of $\chi_R$ arises from the fact that it is
the rescaled $\chi$ that is used to extract $\eta$, rather than from
the variation of $d_s$ itself.

\section{Conclusions}

We have presented a brief addendum to our previous paper
\cite{Drummond:2002yg}
on the one--loop perturbative calculation of the anisotropy in lattice
gauge theories using the on--shell gluon dispersion relation. We have 
demonstrated that the anisotropy renormalisations obtained apply to
actions with arbitrary choice of tadpole improvement scheme. We
re--presented our data to make this explicit.

We pointed out a generalisation of our method. The additive
contributions to the anisotropy of counter--terms in the action of
${\cal O}(\alpha)$ and above may be calculated separately. This is
possible as they do not affect the tree level on--shell condition, and
because the one--loop anisotropy renormalisation coefficient is
related linearly to the first order gluon self energy.

When varying the radiative coefficients, the change in the
renormalisation of the anisotropy is then simply given by a change of
weighting in a sum of previously calculated perturbative contributions
This has obvious benefits during, for instance, the tuning of
radiative corrections to improved lattice actions.

\begin{acknowledgments}

A. Hart is supported by a Royal Society University Research
Fellowship.

\end{acknowledgments}

\end{document}